\documentclass[12pt]{iopart}

\usepackage{graphicx}
\usepackage{xcolor}
\usepackage[urlcolor=blue]{hyperref}      
\hypersetup{
    colorlinks = true,                    
    citecolor = {blue},
    linkcolor = {purple},
           }
\graphicspath{{./}{./figures/}}
\begin{document}

\title{Absorbing phase transitions in a non-conserving sandpile model}

\author{Marvin G\"obel \& Claudius Gros}

\address{Institute for Theoretical Physics, Goethe University Frankfurt,
Frankfurt a.M., Germany}
\ead{gros07[myPlace]@itp.uni-frankfurt.de}
\vspace{10pt}
\begin{indented}
\item[]March 2019
\end{indented}

\begin{abstract}
We introduce and study a non-conserving sandpile 
model, the autonomously adapting sandpile (AAS) model,
for which a site topples whenever it has two or more 
grains, distributing three or two grains randomly on its 
neighboring sites, respectively with probability $p$ 
and $(1-p)$. The toppling process is independent of 
the actual number of grains $z_i$ of the toppling site, 
as long as $z_i\ge2$. 
For a periodic lattice the model evolves into an inactive 
state for small $p$, with the number of active sites 
becoming stationary for larger values of $p$. In one and
two dimensions we find that the absorbing phase transition 
occurs for $p_c\!\approx\!0.717$ and $p_c\!\approx\!0.275$.

The symmetry of bipartite lattices allows states in 
which all active sites are located alternatingly on 
one of the two sublattices, A and B, respectively for
even and odd times. We show that the AB-sublattice 
symmetry is spontaneously broken for the AAS model, 
an observation that holds also for the Manna model. 
One finds that a metastable AB-symmetry conserving 
state is transiently observable and that it has the 
potential to influence the width of the scaling regime, 
in particular in two dimensions.

The AAS model mimics the behavior of integrate-and-fire 
neurons which propagate activity
independently of the input received, as long as the 
threshold is crossed. Abstracting from regular 
lattices, one can identify sites with neurons and consider
quenched networks of neurons connected to a fixed 
number $G$ of other neurons, with $G$ being drawn 
from a suitable distribution. The neuronal activity 
is then propagated to $G$ other neurons. The AAS model 
is hence well suited for theoretical studies of
nearly critical brain dynamics. We also point out that
the waiting-time distribution allows an avalanche-free
experimental access to criticality.
\end{abstract}

%
\vspace{2pc}

%
%
%
%

\section{Introduction}

Dynamical systems with self-propagating
activities are said to undergo an
absorbing phase transition when changes
in a control parameter lead to a transition 
from an active to an inactive, an absorbing 
state \cite{hinrichsen2000non}, or vice versa.
Examples are contact processes \cite{kockelkoren2003absorbing,henkel2004non},
the conserved lattice gas \cite{rossi2000universality},
and sandpile dynamics on lattices with periodic
boundary conditions \cite{manna1991two}.

Absorbing phase transitions are potentially important
for the quasi-stationary neuronal activity of the
brain, which is continuously propagated on 
the level of individual neurons. Indications 
of power-law avalanches
\cite{beggs2003neuronal,priesemann2013neuronal}
have been interpreted in this context as signs of self-organized
criticality \cite{chialvo2004critical}. Power laws
may arise however also from branching processes
that are tuned homeostatically to criticality
\cite{markovic2014power}, namely when the average
activity is self regulated, and when observations
sample dynamical states based on the size of their 
attractors \cite{gros2013observing,markovic2013criticality}, 
as it is the case for vertex routing models \cite{markovic2009vertex}.

Resolving the questions whether the brain is close 
to a self-organized critical state \cite{bak1998nature}, 
as occurring in sandpile models with open boundary 
conditions \cite{bak1987self}, is plagued by several 
issues. On a technical level it has been shown that sub-sampling 
needs to be taken into account when interpreting experiments
\cite{levina2017subsampling}. Conceptually there is
in addition the dichotomy between model and brain with 
respect to conservation laws. Sandpile models with
open boundary conditions are self-organized critical
only when locally conserving
\cite{drossel2000scaling,dickman2003avalanche}, 
at least on the average \cite{vespignani1998driving}, 
a property not shared by real-world neuronal cells.

Absorbing phase transitions in sandpile models with 
periodic boundary conditions have been considered
hitherto to depend on toppling processes that
conserve energy locally \cite{gros2015complex},
or that allow for a fine tuning between dissipation
and energy uptake \cite{basu2013absorbing}. 
Here we propose a new model, the autonomously adapting 
sandpile (AAS) model, for which the toppling events are 
generically non-conserving. Nevertheless, a transition 
between absorbing states and phases with self-regulated
activity is observed. We present a first assessment of
the AAS model, with a focus on several basic issues,
such as the spontaneous breaking of the AB-sublattice 
symmetry and the presence of metastable states.
Our results indicate that the AAS model is likely
to fall into the universality class of directed percolation.

\begin{figure}[t!]
\begin{center}
\includegraphics[width=0.8\textwidth]{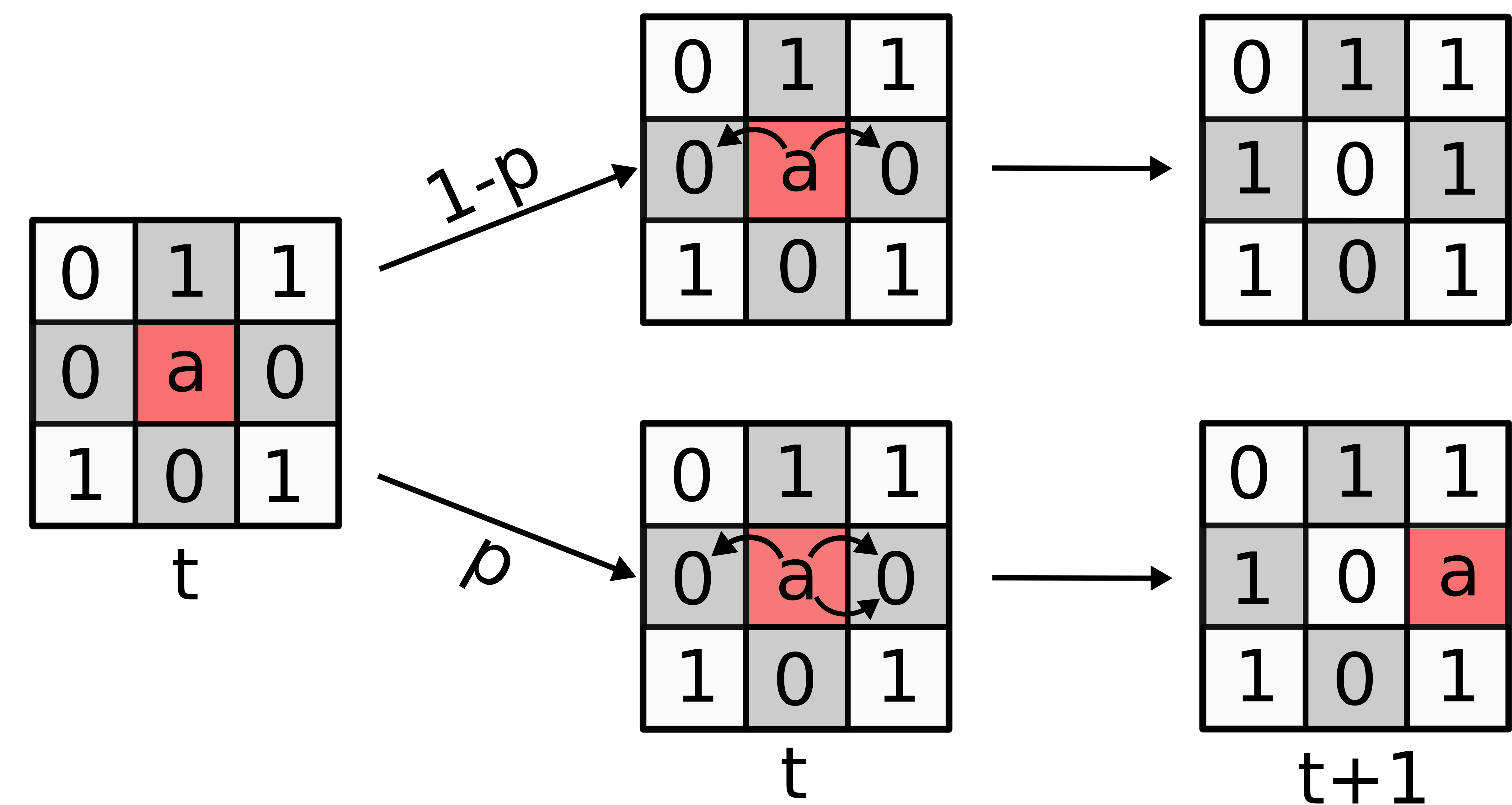}
\end{center}
\caption{Illustration of the AAS toppling rules. 
Active sites (red) become empty after a toppling event, which
consists to adding three or two grains randomly to 
neighboring sites, respectively  with probabilities $p$ and $(1-p)$. 
Note that the number of grains added is independent of the 
occupation number $z_i$ of the toppling site, as long as 
$z_i\ge2$. Sites may receive more than one grain.
}
\label{AAS_topplingRules}
\end{figure}

\subsection{The autonomously adapting sandpile model}

\begin{figure}[t!]
\begin{center}
\includegraphics[width=0.90\textwidth]{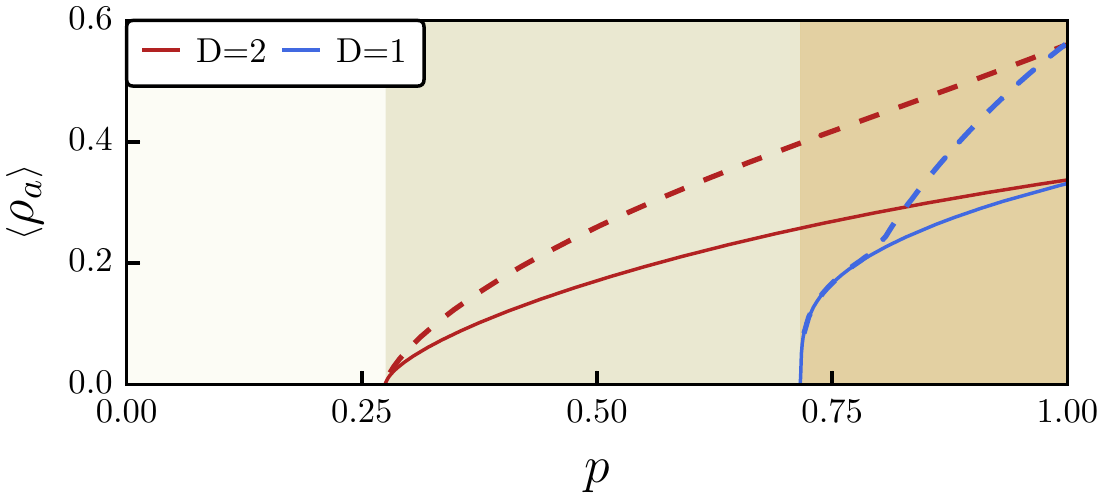}
\end{center}
\caption{The density $\rho_a$ of active sites, as a
function of the probability $p$ that a toppling event 
distributes three and not two grains, as illustrated in 
Fig.~\ref{AAS_topplingRules}. For $L\!\times\!L$ 
square lattices with $L\!=\!128\!-\!512$ (red lines), and for a one-dimensional 
system with $10^4\!-\!10^5$ sites (blue lines). Synchronous 
updating has been used, which allows for metastable 
solution conserving the AB-sublattice symmetry (dashed lines),
as discussed in Sect.~\ref{sect_AB}. After prolonged times
the system settles into a stable AB-sublattice symmetry 
breaking solution (full lines). Close to the absorbing 
transition the metastable solution is not well defined, 
as evident is particular in one dimension. 
The transitions from an absorbing to an active phase occur 
at $p_c\!\approx\!0.27453(1)$ and $p_c\!\approx\!0.71692(2)$ 
in two and one dimensions.
}
\label{phaseDiagram}
\end{figure}

The toppling dynamics of the AAS model is
a straightforward modification of the respective
Manna model rules. A site with two or more grains
topples, at which point either two or three grains
are added randomly onto its neighboring sites,
as illustrated in Fig.~\ref{AAS_topplingRules}.
The key difference to the Manna model is that
the number of grains added to the neighboring sites 
is independent of the number $z_i$ of grains 
on the toppling site before toppling.
Instead, three grains are distributed with probability
$p$ and two grains with probability $1-p$. After toppling
the formerly active site has zero grains associated
with it. As shown in Fig.~\ref{phaseDiagram}, one observes
an absorbing phase transition, which occurs 
at $p_c=0.27523(1)$ in two dimensions and
at $p_c=0.71692(2)$ in one dimension.

The mechanism leading to a self-organized activity
level in the AAS model is closely related to the 
homeostatic adaption regulating the firing rate of 
real-world neurons. A site is active when the
number of grains received is larger than a certain
threshold, which is unity in the here considered case.
The number of grains and with it the activity is 
down-regulated when $z_i$ is large, as no more than 
three grains can be distributed. The activity is on 
the other hand up-regulated when $z_i$ is small, 
in our case when $z_i=2$, as there is a finite 
probability $p$ to distribute more grains, namely three.

Within the AAS model discussed here, either
two or three grains are distributed. Generalizing
this rule one can define a distribution $p(G)$, such
that $G>0$ grains are distributed during a toppling
event with probability $p(G)$. Instead of a regular 
lattice one can consider in addition a network of 
sites having a coordination number, the average number 
of edges \cite{gros2015complex}, that may be either 
small or large. We believe that the resulting model, 
the net-AAS, would capture key features of interconnected
integrate-and-fire neurons and that it would be interesting 
to use the net-AAS model for theoretical studies of critical 
neuronal activity. 

Of experimental interest are furthermore methods that 
allow to gauge the distance to criticality without the 
need to define avalanches. We point out that the distribution
of waiting times between two topplings events is critical 
and easily measurable, as it corresponds to the time between 
subsequent neuronal spikes. For the AAS model we confirm 
that the mean waiting time scales inversely to the number 
of active sites \cite{henkel2008non}, viz to the order parameter. 

\begin{figure}[t!]
\begin{center}
\includegraphics[width=0.9\textwidth]{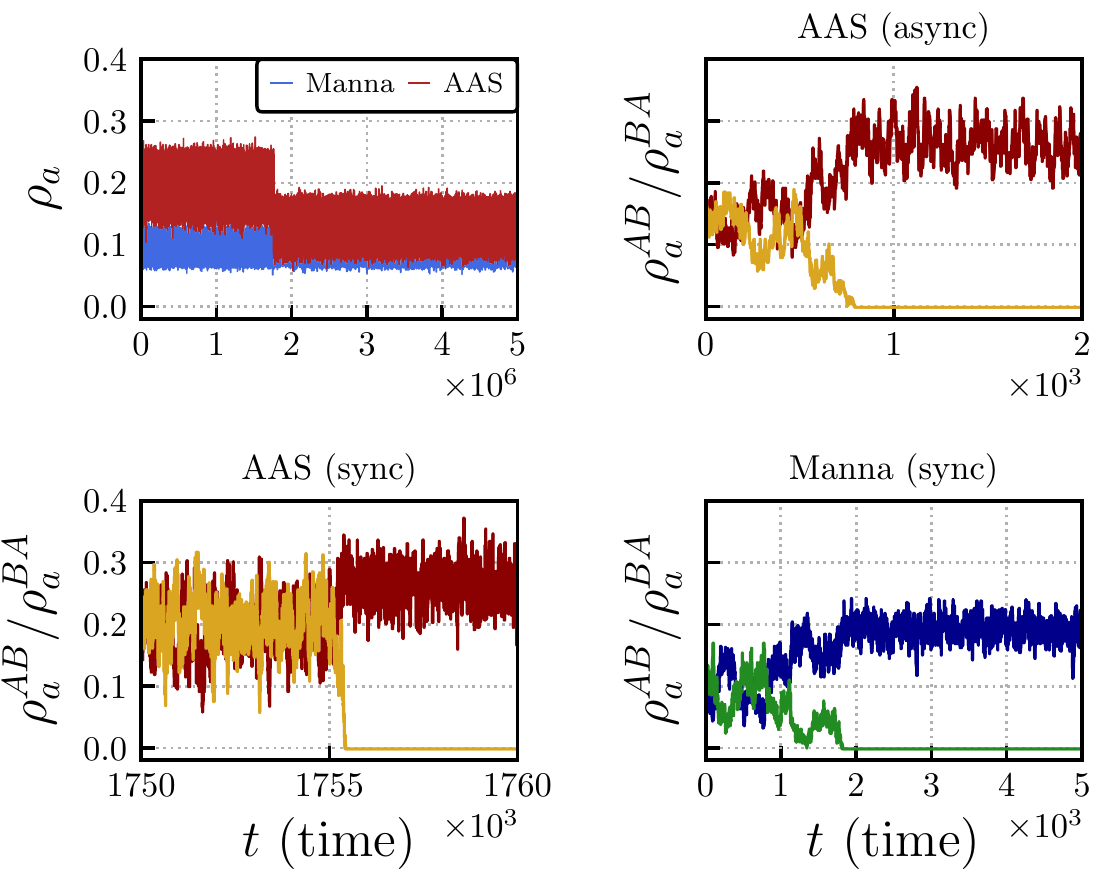}
\end{center}
\caption{Timelines illustrating the spontaneous breaking of the 
AB-sublattice symmetry on a $32\!\times\!32$ square lattice, for
the AAS model with $p\!=\!0.4$ (red and yellow curves) and for the 
Manna model with a grain density of $0.74$ (blue and green curves). 
Shown are results for synchronous (sync) and sequential
asynchronous (async) updating.
{\em Upper Left:} The overall density of active sites, which
remains finite, as both systems are in the active phase. Evident 
is a transition, which occurs after about $1.76\cdot10^6$ steps
for the AAS model. At this scale the Manna model is already 
in a symmetry broken state. 
{\em Upper Right:} A close-up of the densities of active sites,
$\rho_a^{\mathrm{AB}}$ and $\rho_a^{\mathrm{BA}}$, which alternate
between the two sublattices for even/odd times. AAS model 
with asynchronous updating. 
{\em Lower Left:} $\rho_a^{\mathrm{AB}/\mathrm{BA}}$ for the
AAS model with synchronous updating. 
{\em Lower Right:} $\rho_a^{\mathrm{AB}/\mathrm{BA}}$ for the
Manna model with synchronous updating. 
}
\label{sublatticeTransients_2D}
\end{figure}

\begin{figure}[t]
\centering
\includegraphics[width=0.9\textwidth]{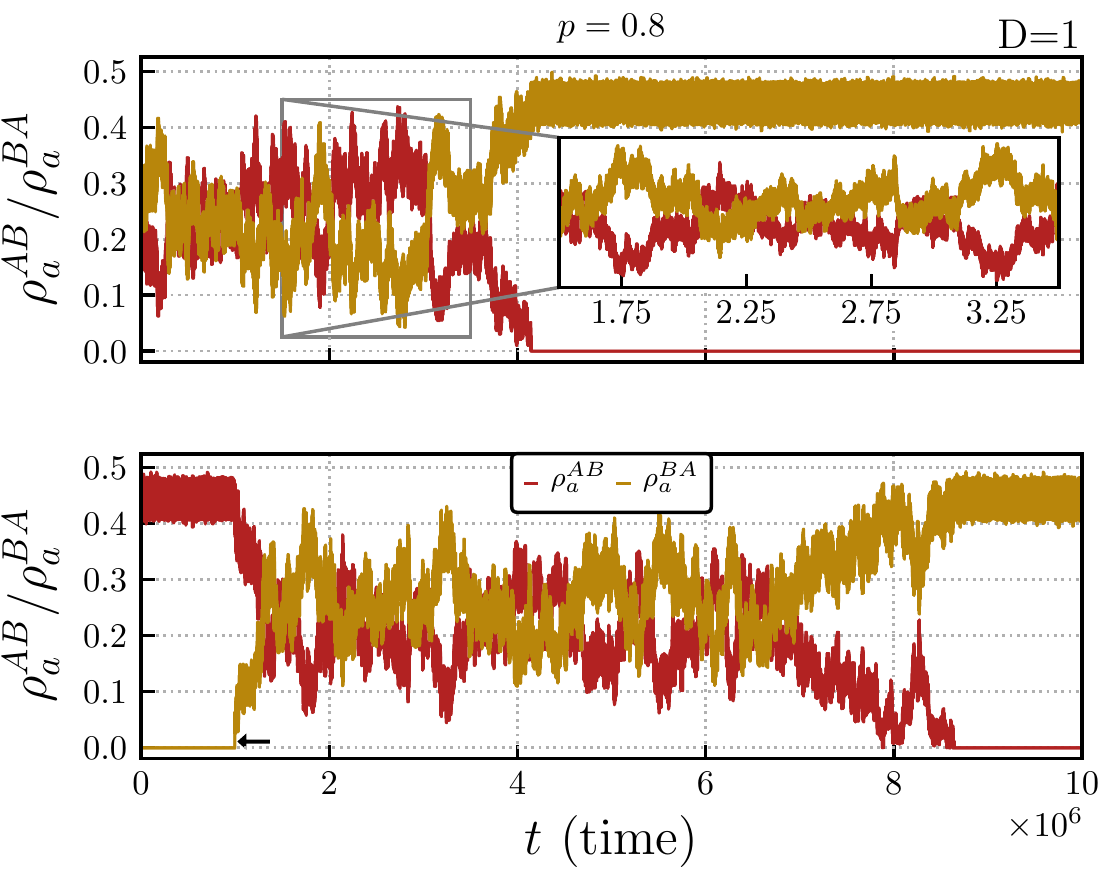}
\caption{For $p\!=\!0.8$ the time series of the densities 
of active sites $\rho_a^{\mathrm{AB}}/\rho_a^{\mathrm{BA}}$ 
located alternating for even and odd times on the
$\mathrm{A}$ and $\mathrm{B}$ sublattice 
in the one-dimensional AAS model with $L\!=\!10^4$. 
Below $p\!\approx\!0.785$ the AB-conserving and broken
states are nearly indistinguishable, compare 
Fig.~\ref{phaseDiagram}.
\textit{Upper panel:}
The symmetry is broken spontaneously after about $4\cdot10^6$ 
time steps. The symmetric state is characterized by visible 
drifts in $\rho_a^{\mathrm{AB}}/\rho_a^{\mathrm{BA}}$ whereas 
only small drifts can be observed in the asymmetric 
state. 
\textit{Lower panel:} 
The system was initialized with a broken AB-symmetry. At $t\!=\!10^6$ 
the grains of $5\%$ of the active sites were shifted by one 
lattice spacing in a random direction, leaving the total number
of grains and active sites conserved. This procedure generates 
activity on the formerly inactive sublattice, as indicated by the
horizontal arrow. For short times the symmetric metastable state is
restored until the sublattice symmetry is broken anew by 
stochastic fluctuations.
}
\label{sublatticeTransients_1D}
\end{figure}

\section{Stochastic breaking of the AB-sublattice symmetry\label{sect_AB}}

As bipartite lattices, one-dimensional chains and 
the square lattice decompose into two sublattices, A and 
B. Active sites on the A sublattice 
distribute grains exclusively to the B sublattice, 
and vice versa. The system may hence settle into a 
sequence of configurations for which all active sites 
are located alternatively on the A and the B sublattice. 
Such a sequence would imply that the AB-sublattice symmetry 
is dynamically broken. Whether this is possible depends 
however on the updating rule.
\begin{itemize}
\item {\em Synchronous updating} consists of saving the
      current configuration in a first step, with the
      subsequent toppling of the active sites occurring
      simultaneously \cite{hinrichsen2000non}.
\item For {\em sequential asynchronous updating} one
      goes randomly through the list of active sites.
      Once processed, a new list of active sites is compiled 
      \cite{lubeck2001scaling}.
\item {\em Random asynchronous updating} implies in contrast
      that active sites are toppled randomly one by one.
      The configuration is updated after every toppling
      event \cite{lee2014critical}.
\end{itemize}
It is clear that random asynchronous updating does not 
allow the AB-sublattice symmetry to be broken dynamically.
For synchronous and sequential asynchronous dynamics
one observes however, as shown in Fig.~\ref{sublatticeTransients_2D}, 
that the AB-sublattice symmetry is spontaneously broken
for the AAS and the Manna model. 
These results indicate that the AB-sublattice symmetry 
may be generically broken for sandpile models 
with nearest-neighbor dynamics on bipartite lattices.

We denote with $\rho_a^{AB}(t)$ the density of active 
sites on the A sublattice, when $t$ is even, and, alternatingly,
the density of active sites on the B sublattice 
for odd times $t$. An equivalent definition holds for
$\rho_a^{BA}(t)$. Starting from random initial conditions, 
for which $\rho_a^{AB}(t)$ and $\rho_a^{BA}(t)$ are of 
comparable magnitude, one observes that $\rho_a^{AB/BA}(t)$
fluctuate until one of the sublattices is devoid of
active sites, compare Fig.~\ref{sublatticeTransients_2D}. 
At this point the AB-sublattice symmetry is fully broken.

Empirically we find that the time $t_0$ needed for
breaking the AB-sublattice symmetry increases
with system size. We did however restrain from
attempting to determine the functional dependency 
of $t_0$ on the system size and on $p-p_c$, noting 
that $t_0$ tends to vary strongly from one simulation 
to another.
We could hence not determine, at this point, whether 
the process leading to the breaking of the AB-sublattice 
symmetry shows finite-size and/or critical scaling.
It is presently also not clear why $t_0$ differs by 
several orders of magnitude between the Manna and the 
AAS model, and/or for synchronous vs.\ sequential
asynchronous updating. This disparity indicates in 
any case that the AB-symmetry is broken via a 
stochastic process and not by a driving molecular 
field, viz already on a molecular-field level, a 
supposition that is in agreement with the 
mean-field results presented in Sect.~\ref{sect_MFT_AB}.

\begin{figure}[t]
\centering
\includegraphics[width=0.90\textwidth]{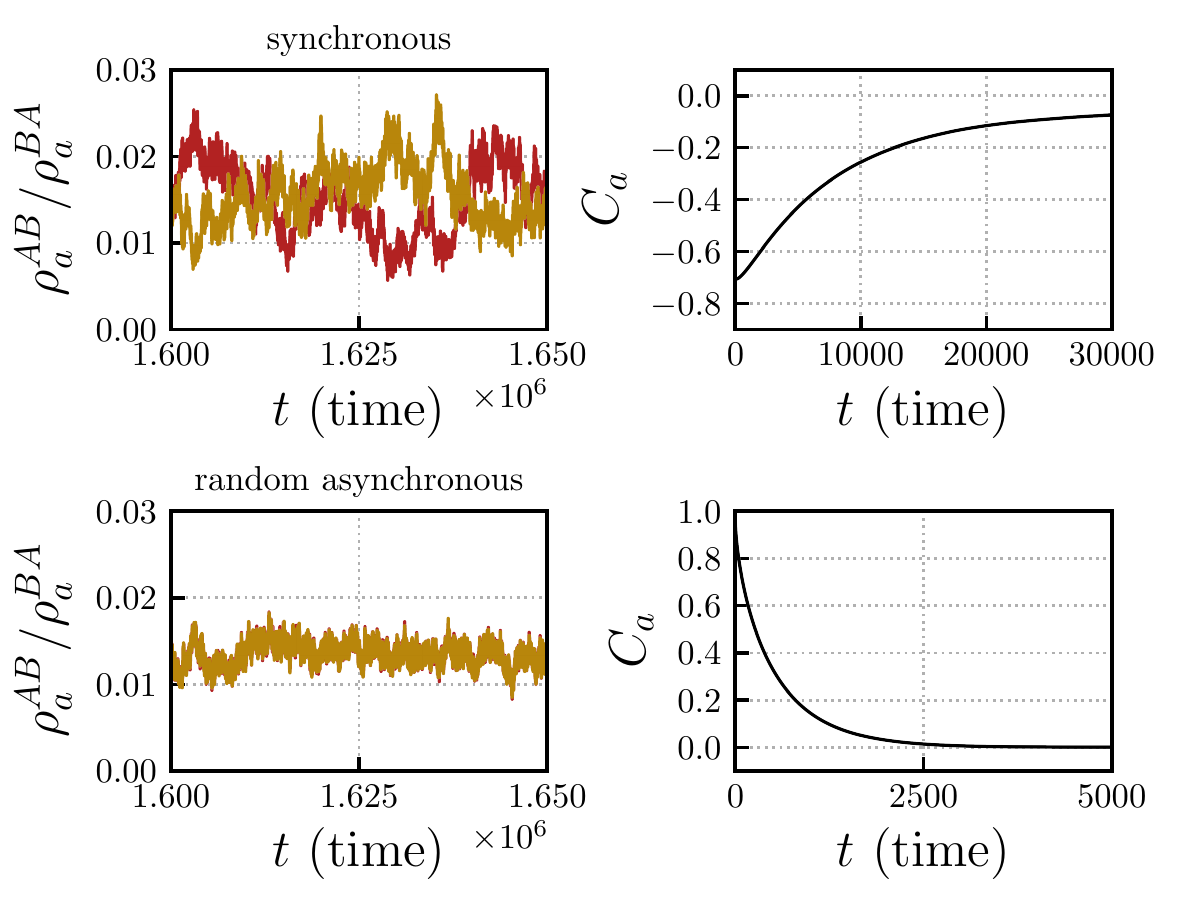}
\caption{
The timeline of the number of active sites located alternatingly
on the A and B sublattice, $\rho_a^{\mathrm{AB}}$ and
$\rho_a^{\mathrm{BA}}$. For a $512\!\times\!512$ square lattice 
with synchronous updating (upper panels) and with
random asynchronous updating (lower panel). The AB-sublattice
symmetry is not broken in the latter case. Here
$p\!-\!p_c\!=\!4\!\cdot\!10^{-3}$. Also shown are the
respective cross-correlation functions $C_a$, as
defined by (\ref{def_C_a}). Note that the sublattices
are highly anti-correlated for the metastable state
realized for synchronous updating.
}
\label{fig:2D_breaking_random}
\end{figure}

\subsection{Metastable state}

The long timescales needed to break the AB-sublattice
symmetry for synchronous update rules allow to study
the metastable sublattice symmetric state. 
In Fig.~\ref{sublatticeTransients_1D} we present 
the evolution for the one-dimensional case, which 
indicates that stochastic tipping may cause the
instability of the AB-symmetric state. For large
densities of active sites the instability cannot
be observed, presumably due to diverging time scales.

We note, as shown in Fig.~\ref{phaseDiagram}, 
that the densities of active sites of the two states, 
AB-sublattice conserving and not conserving, approach
each other for $p\to p_c$. In 1D the two densities
become virtually indistinguishable below 
$p\approx 0.785$, viz substantially above $p_c$.
The stability time of the metastable state is however
reduced when the densities of active sites of the 
two states become similar.

Also included in Fig.~\ref{sublatticeTransients_1D} 
is a protocol studying the short-term stability 
of the symmetry-broken state. Inducing a small number 
of active sites on the sublattice that was previously 
devoid of active sites leads to a short-term runaway 
growth, until the fluctuating sublattice-symmetric 
metastable state is reached. This implies that
the asymmetric state is unstable on short time
scales, but absorbing for long times, which 
is possibly a quite unique situation. 

A symmetry can be broken stochastically only if the
broken state is dynamically stable, viz absorbing.
A stochastically broken symmetry could hence influence
the universality class \cite{hohenberg1977theory}. 
It seems however, that the exponents found numerically
for models that are prone to break the AB-sublattice 
symmetry are independent of the update rules used 
\cite{lubeck2003universal,daga2018universality}.
We did not investigate these issues further, concentrating
in this study for the most part on the ASS model 
with synchronous updating.

\subsection{Correlations in the metastable state}

In order to obtain an improved understanding of the
metastable state we show in Fig.~\ref{fig:2D_breaking_random}
a typical timeline for a $512\times512$ two-dimensional 
system close to the transition point, specifically for
$p-p_c=4\cdot10^{-3}$. The densities of active sites 
on alternating sublattices, $\rho_a^{\mathrm{AB}}$ and
$\rho_a^{\mathrm{BA}}$, are manifestly anticorrelated.
For a quantitative measure we evaluated the
time-dependent cross-correlation function 
\begin{equation}
C_a(t) = \frac{\biggr\langle \Big(\rho_a^{AB}(t) - \langle \rho_a^{AB} \rangle \Big) 
\Big( \rho_a^{BA}(0) - \langle \rho_a^{BA} \rangle \Big) \biggr\rangle}
{\sigma^{AB} \cdot \sigma^{BA}}\,,
\label{def_C_a}
\end{equation}
where
\begin{equation}
\sigma^{AB} = \sqrt{\Big \langle \left(\rho_a^{AB} - 
\langle \rho_a^{AB} \rangle \right)^2 \Big\rangle}, 
\qquad 
\sigma^{BA} = \sqrt{\Big \langle \left(\rho_a^{BA} - 
\langle \rho_a^{BA} \rangle \right)^2 \Big\rangle}\,.
\label{def_sigma_AB}
\end{equation}
One observes that $C_a$ is always negative, starting
out at about $-0.7$, which implies a substantial 
anticorrelation of the two sublattices.

Included in Fig.~\ref{fig:2D_breaking_random} are the 
results for a simulation using random asynchronous updating, 
for which toppling events are treated individually, as 
explained further above. For random asynchronous dynamics
the sublattice symmetry can not be broken. Interestingly, 
the cross-correlation function $C_a$ is in this case always
positive, decaying furthermore substantially faster than 
for synchronous updating. The respective times to decay
to half-heights are about 6000 and 300, respectively
for synchronous and random updating. Note that the time 
unit is given in the synchronous by a sweep and in the 
asynchronous by $N_a$ topping events, where $N_a$ is 
the current number of active sites, which implies that the 
two time units are comparable. It would be interesting, but
beyond the scope of the present investigation, to study 
the metastable state in more detail, as well as the
properties of systems with random asynchronous updating.

\begin{table}[b]
\caption{One-dimensional critical exponents. Note
that $\tilde{\beta}$, $\tilde{\gamma}'$ and $\tilde{\sigma}$ 
are defined only for the autonomously adapting sandpile 
model (AAS), see (\ref{active_grain_beta_tilde}) and
(\ref{Delta_active_grain_exp})
and (\ref{field_exp}). Manna and directed 
percolation (DP) exponents are from \cite{lubeck2004universal}. 
The exponent $\epsilon_\parallel/\nu_\parallel$ for the
waiting time distribution is defined
by (\ref{P_wait_epsilon_nu}).
\label{tab_exponents_1D}
}
\centering
\begin{tabular}{llll}
1D & AAS & Manna & DP \\
\hline \hline
$ \beta$           & 0.278(5)  & 0.382(19)     & 0.276486(8)  \\
$ \tilde{\beta}$   & 0.293(14) & -             & -            \\
$ \gamma'$         & 0.54(3)   & 0.55(4)       & 0.543882(16) \\
$ \tilde{\gamma}'$ & 0.52(3)   & -             & -            \\
$ \sigma$          & 2.40(8)   & 2.71(4)       & 2.554216(13) \\
$ \tilde{\sigma}$  & 2.46(9)   & -             & -            \\
\hline
$\epsilon_\parallel/\nu_\parallel$
                   & 1.86(9)   & -             & 1.840536(5)  \\
\end{tabular}
\end{table}

\section{Critical properties}

Numerical simulations of the AAS model are
somewhat demanding, as the number of grains 
fluctuates in addition to the number of active sites.
We performed simulations in one dimension
and for $L\times L$ square lattices, using synchronous 
updating, which defines a time step as a sweep over 
all active sites. Typically we reserved $5\cdot10^7$ 
simulation steps for equilibration, with the actual 
measurements taking about $10^8$ time steps, which
reflects the procedure used in comparable studies
\cite{basu2013absorbing,basu2012fixed}.
We checked that the initial grain density is irrelevant,
viz that the steady-state density $g=g(p)$ is uniquely
a function of $p$. No differences with respect to 
synchronous updating could be found when
performing somewhat less extensive simulations with 
sequential and random asynchronous updating rules, 
as defined in Sect.~\ref{sect_AB}.
For the correlation length, we found that the numerical 
efforts required to obtain reliable results are exceedingly 
demanding. This topic was hence left to future studies.

For synchronous updating it may take a substantial time 
for the AB-sublattice symmetry to be broken
stochastically, as described in Sect.~\ref{sect_AB}.
In order to make sure that we work in the
symmetry broken state we used symmetry breaking
initial conditions, with the overall density of grains
being one. Grains were added one by one to a site
either if the site was on the A sublattice, or if it
was empty.

\subsection{Scaling exponents}

For the densities $\langle\rho_a\rangle$ 
and $\langle g\rangle$ of active sites and
respectively of grains, the scaling exponents
are defined as
\begin{equation}
\langle\rho_a\rangle\sim (p-p_c)^\beta,
\qquad\quad
\langle g\rangle-g_c\sim (p-p_c)^{\tilde{\beta}}\,,
\label{active_grain_beta_tilde}
\end{equation}
where we have used with $\beta$ and $\tilde{\beta}$
a notation that relates the exponents of the grain 
density to the ones characterizing the scaling of 
the density of active sites. The time averaged 
fluctuations of the density of active sites 
and grains,
\begin{equation}
\Delta\rho_a = L^2\big\langle (\rho_a-\langle\rho_a\rangle)^2
                  \big\rangle,
\qquad\quad
\Delta g     = L^2\big\langle (g     -\langle g    \rangle)^2
                  \big\rangle\,,
\label{Delta_active_grain}
\end{equation}
come with respective exponents,
\begin{equation}
\Delta\rho_a \sim (p-p_c)^{-\gamma'},
\qquad\quad
\Delta g  \sim (p-p_c)^{-\tilde{\gamma}'}\,.
\label{Delta_active_grain_exp}
\end{equation}
We also evaluated the mean time 
$\langle \tau_{\mathrm{wait}}\rangle$
between two toppling events on the same site, 
the waiting time,
\begin{equation}
\qquad\quad
\langle \tau_{\mathrm{wait}}\rangle =1/\langle\rho_a\rangle 
\sim (p-p_c)^{-\beta}\,,
\label{tau_wait}
\end{equation}
where we have included that the waiting
time is inversely proportional to the density of
active sites, compare (\ref{Delta_active_grain}).
Within the neural network interpretation the
mean waiting time corresponds to the average
interspike interval, which is experimentally 
accessible. Measuring $\langle \tau_{\mathrm{wait}}\rangle$
allows hence to estimate $\beta$ and to gauge 
whether the scaling regime is attained.

For estimating the accuracy of the critical exponents, 
we subsampled the data to which the fit was performed. 
Typically, we took $10^3$ sets containing each six 
randomly chosen data points from the selected region, 
to which power laws where fitted. The error of the 
respective exponents was then taken as an estimate 
of the final standard deviation \cite{HINRICHSEN20061}.

\begin{figure}[t!]
\begin{center}
\includegraphics[width=0.90\textwidth]{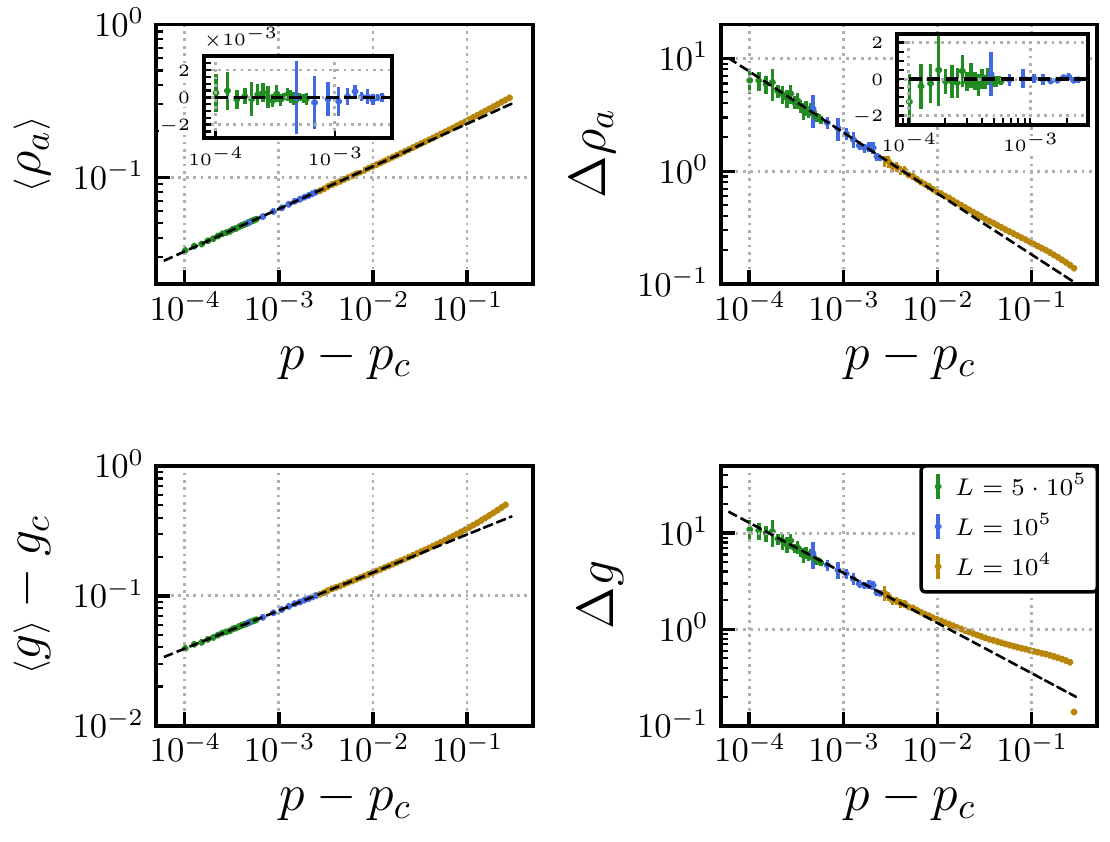}
\end{center}
\caption{For one-dimensional chains of length $L$, 
the moments of the density of active sites and grains 
in the AB-symmetry broken state. In the insets for 
$\langle\rho_a\rangle$ and $\Delta\rho_a$ the error
relative to the respective power-law fit is shown in
units of $10^{-3}$ and $10^0$.
The fits for $\langle\rho_a\rangle$ and $\langle g\rangle-g_c$
where performed simultaneously, assuming a common critical probability
to distribute three and not two sites, $p_c\!=\!0.71692(2)$, upon toppling. 
The resulting critical density of grains is $g_c\!=\!0.733(6)$.
Data for up to $p\!=\!1$ is shown.
\textit{Upper panels:} The density and the fluctuations of active sites 
as a function of $p-p_c$ (colored symbols), in a log-log representation,
together with the optimal fit (dashed line). The exponents are
$\beta\!=\!0.278(5)$ and $\gamma'\!=\!0.54(3)$. 
\textit{Lower panels:} 
As for the upper panels, but for the mean grain density 
$\langle g\rangle$. 
The exponents are $\tilde{\beta}\!=\!0.293(14)$ and 
$\tilde{\gamma}'\!=\!0.52(3)$.
}
\label{AAS_active_g_Delta_1D}
\end{figure}

\subsection{1D scaling}

The numerical results for the scaling of the
densities of active sites and grains are presented
in Fig.~\ref{AAS_active_g_Delta_1D} for the case
of chains with a variable number of $L$ sites.
We find that the scaling exponents for the
density of active sites and for the
density of grains are distinct, but very close,
$\beta=0.278(5)$ and $\tilde{\beta}=0.293(14)$, see
(\ref{active_grain_beta_tilde}). The same observation
holds for the scaling exponents of the corresponding
fluctuations, which are $\gamma'=0.54(3)$ and 
$\tilde{\gamma}'=0.52(3)$. Our resolution allows 
for identical scaling exponents for the density 
active sites and grains, a result that is
consistent with the observation that the density 
of active sites and the mean energy per site show 
similar scaling behavior for a two-parameter 
non-conserving sandpile model \cite{basu2013absorbing}.

In Table~\ref{tab_exponents_1D} the critical exponents
obtained for the 1D autonomously adapting sandpile model
are listed together with the estimates for the
Manna model and directed percolation (DP). The results
indicate that the universality class of the AAS model 
is DP, in agreement with the conjecture that generic 
sandpile models have directed percolation exponents 
\cite{mohanty2002generic}.

The range of $p$ included in Fig.~\ref{AAS_active_g_Delta_1D} 
reaches $p=1$ at the upper end for the probability to
distribute three and not two grains. Deviations from 
a power law start to show up at around a distance of 
$p-p_c=\Delta p\approx\!10^{-2}$ from the critical point, 
in particular for the fluctuations
$\Delta g$ of the grain density.

\begin{figure}[t!]
\begin{center}
\includegraphics[width=0.90\textwidth]{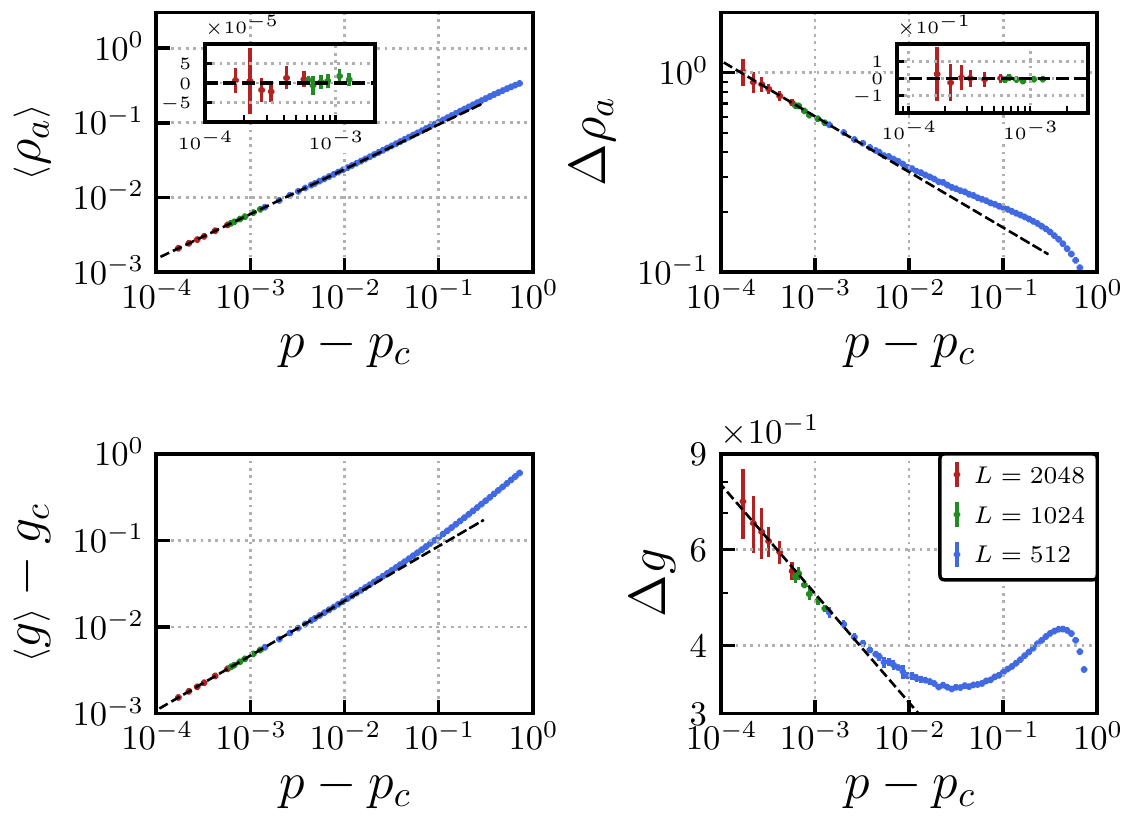}
\end{center}
\caption{For two dimensional $L\times L$ square lattices, the 
scaling behavior as a function of the probability $p$ for 
toppling three and not two sites, with the data reaching
to $p=1$. Dashed lines are power-law fits.
{\em Top left:} The density of active sites $\rho_a$
fitted by $\rho_a\sim (p-p_c)^\beta$, with
$p_{c}=0.27523(1)$ and $\beta=0.599(9)$.
The inset shows the error relative to the fit,
in units of $10^{-5}$.
{\em Top right:} The fluctuations $\Delta\rho_a$ of 
active sites, with an exponent $\gamma'=0.278(12)$.
{\em Bottom left:} The relative density of grains 
$\langle g\rangle-g_c$ with an exponent
$\tilde{\beta}=0.632(9)$ and $g_c=0.64686(4)$.
The inset shows the error relative to the fit,
in units of $10^{-1}$.
{\em Bottom right:} The fluctuations $\Delta g$ 
of the density of grains with an exponent 
$\tilde{\gamma}'=0.2$. The quality of the data
prevented us to obtain reliable estimates
for the accuracy of $\tilde{\gamma}'$.
}
\label{AAS_active_g_Delta_2D}
\end{figure}

\begin{table}[b]
\caption{Two-dimensional critical exponents. The scaling
exponents $\tilde{\beta}$, $\tilde{\gamma}'$ and $\tilde{\sigma}$ 
for the grain density are defined only for the autonomously 
adapting sandpile model (AAS), see (\ref{active_grain_beta_tilde}) 
and (\ref{Delta_active_grain_exp}) and (\ref{field_exp}).  
Manna and directed percolation (DP) exponents are from 
\cite{lubeck2004universal}. 
For the definition of $\epsilon_\parallel/\nu_\parallel$ see
(\ref{P_wait_epsilon_nu}). The poor scaling behavior of
the numerical data prevented us from obtaining reliable
estimates for the accuracy of $\tilde{\gamma}'$
and $\tilde{\sigma}$.
\label{tab_exponents_2D}}
\centering
\begin{tabular}{llll}
2D & AAS & Manna & DP \\
\hline \hline
$ \beta$           & 0.599(9)   & 0.639(9)  & 0.5834(30)  \\
$ \tilde{\beta}$   & 0.632(9)       & -         & -           \\

$ \gamma'$         & 0.278(12)  & 0.367(19) & 0.2998(162) \\
$ \tilde{\gamma}'$ & 0.2        & -         & -           \\

$ \sigma$          & 1.99(4)    & 2.229(32) & 2.1782(171) \\
$ \tilde{\sigma}$  & 1.88       & -         & -           \\
\hline
$\epsilon_\parallel/\nu_\parallel$
                   & 1.51(11)   & -         & 1.549(3)    \\
\end{tabular}

\end{table}

\begin{figure}[t]
\begin{center}
\includegraphics[width=0.90\textwidth]{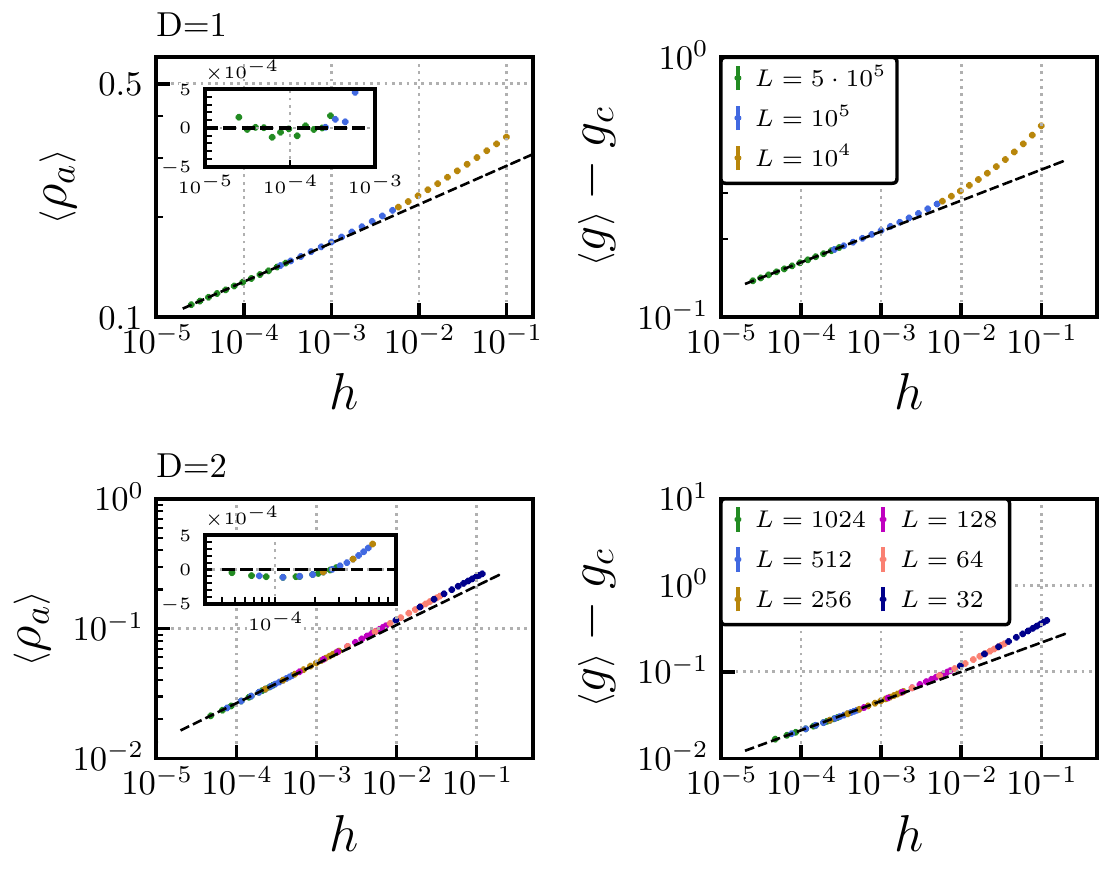}
\end{center}
\caption{The density of active sites and grains as a function of 
the external field $h$ and for $p\approx p_c$. The insets show
the deviation from the respective power-law fits, in units of $10^{-4}$.
\textit{Upper panels:} In 1D, for chains of length $L$.
The scaling exponents are $\beta/\sigma=0.116(3)$ 
and $\tilde{\beta}/\tilde{\sigma} = 0.119(4)$ (dashed lines). Compare
Sect.~\ref{Sect_field} and (\ref{field_exp}).
\textit{Lower panels:} In 2D, for $L\times L$ lattices. 
The power laws for the order parameter and the grain 
density have exponent $\beta/\sigma= 0.301(3)$
and $\tilde{\beta}/\tilde{\sigma}=0.338(12)$.
}
\label{AAS_1D_2D_field}
\end{figure}

\subsection{2D scaling}

In Fig.~\ref{AAS_active_g_Delta_2D} the results
for the mean density of of active sites and
grains are presented together with their respective
fluctuations. For a comparison we have included
data for distribution parameters $p$ that
extend up to $p=1$. The fluctuations,
$\Delta\rho_a$ and $\Delta g$ start to deviate from
a power-law scaling already for $p-p_c$ substantially
below $10^{-2}$, which is in particular evident
for $\Delta g$. This pronounced deviation from a 
power-law scaling, which may be due to the presence 
of the metastable state, as discussed in 
Sect.~\ref{sect_metatstable}, prevented us to
obtain an reliable estimate for the scaling of $\Delta g$. An error 
bar for the respective exponent could likewise 
not be evaluated. The differences found between 
$\gamma'$ and $\tilde{\gamma}'$, as evident from
Table \ref{tab_exponents_2D}, may consequently
be due to the difficulty to obtain reliable
estimates in particular for $\tilde{\gamma}'$.
Regarding $\beta$ and $\tilde{\beta}$, our
results indicate that they may be distinct
in two dimensions, but not in 1D, see 
Table \ref{tab_exponents_1D}. At this point
we cannot settle definitively whether 
$\beta$ and $\tilde{\beta}$ are identical or
not.

For the Manna model the grain density $g^{Manna}$
is the control parameter, and not a self-consistently
determined quantity. All scaling relations are evaluated
hence as a function of $g^{Manna}-g_c^{Manna}$, 
where $g_c^{Manna}$ is the critical grain density,
estimated to be $0.68333(3)$ \cite{lubeck2004universal},
or $0.68354(1)$ \cite{lee2014critical}. These estimates
for the two-dimensional Manna model differ in any case 
from the 2D-AAS value $g_c=0.64686(4)$. 

The scaling of $\langle\rho_a\rangle$ as a function of
$g^{Manna}-g_c^{Manna}$ defines for the Manna model 
the order-parameter exponent $\beta$. For the 
two-dimensional Manna model $\beta$ has been estimated 
to be $0.639(9)$ \cite{lubeck2004universal} and
$0.634(3)$ \cite{lee2014critical}, with both values
being compatible with $\tilde{\beta}=0.632(9)$, as
defined by (\ref{active_grain_beta_tilde}). Our 
value for $\beta=0.599(9)$ is on the other side 
compatible with the DP estimate $0.5834(30)$ 
\cite{lee2014critical}.

\subsection{External field\label{Sect_field}}

There are several ways to define an external field
$h$ for sandpile models. For conserving models 
in $D$ dimensions one usually transfers $L^Dh$ 
grains between sites, which then results in the 
creation of active sites \cite{lubeck2002scaling}. 
For directed percolation models one generates directly
active sites \cite{lubeck2002universal}. 

The number of grains is not conserved for the 
autonomously adapting sandpile model, a circumstance 
that allows a straightforward definition of a field 
that is conjugated to the order parameter. For this purpose 
$L^Dh$ grains are added randomly onto the active 
sublattice, where $h$ is the field strength. This 
is done at the start of each time step, viz of every 
sweep. The density of active sites and grains then scale as
\begin{equation}
\langle\rho_a\rangle\sim h^{\beta/\sigma},
\qquad\quad
\langle g\rangle-g_c\sim h^{\tilde{\beta}/\tilde{\sigma}}
\label{field_exp}
\end{equation}
at criticality. An external field corresponding to
an additional flux of grains arises naturally
within the neural network interpretation of the AAS 
model. The field is then equivalent to the external 
input of an otherwise recurrent network. Formally
we defined in (\ref{field_exp}) exponents $\sigma$ 
and $\tilde{\sigma}$, which characterize both the
scaling of the field. One hence expects that
$\sigma=\tilde{\sigma}$.

In Fig.~\ref{AAS_1D_2D_field} we present the results
for both 1D and 2D systems. Shown are the field
dependencies of $\langle\rho_a\rangle$ and $\langle g\rangle-g_c$
close to criticality, viz for $p=p_c$, where $p_c$
is the numerically determined critical probability to
topple three and not two grains. The resulting
scaling exponents $\sigma$ and $\tilde{\sigma}$
are somewhat close to the DP results, as listed
in the Tables \ref{tab_exponents_1D} and 
\ref{tab_exponents_2D}. The difficulty to
determine $p_c$ with higher accuracy may contribute to
the remaining difference.

\begin{figure}[t]
\centering
\includegraphics[width=0.90\textwidth]{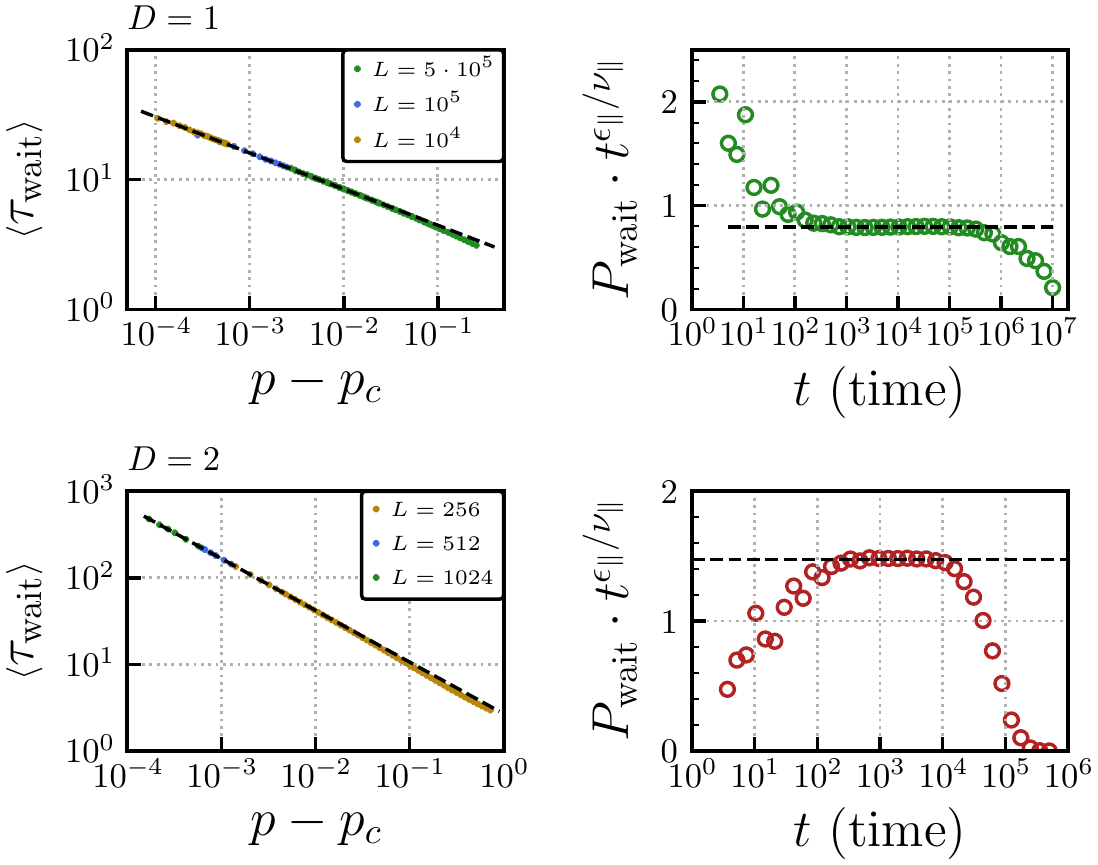}
\caption{
The waiting time and its distribution in 1D
(upper panels) and in 2D (lower panels).
\textit{Left panel:} The mean waiting time plotted against 
the distance $p-p_c$ from the critical point. A regression 
analysis yields $-0.277(10)$ and $-0.597(10)$ for the exponent 
of the mean waiting time in 1D and 2D, which agrees with the 
respective values of the order-parameter exponent $\beta$, see 
the Tables \ref{tab_exponents_1D} and \ref{tab_exponents_2D}.
\textit{Right panels:} For even times, the rescaled waiting time 
distribution in log-log representation. The data is
for $p-p_c=1.05\cdot10^{-4}$ and $L=5\cdot 10^5$
in 1D and for $p-p_c=1.7\cdot10^{-4}$ and $L=2048$
in 2D. A regression analysis (dashed lines) yields 
an exponent $\epsilon_{\parallel}/\nu_\parallel$, as 
defined by (\ref{P_wait_epsilon_nu}), of $1.86(9)/1.51(11)$ 
in 1D/2D. 
}
\label{fig:AAS_Waiting_1D_2D}
\end{figure}

\subsection{Waiting-time distribution}

The waiting-time distribution, which is also known 
as the temporal empty-interval distribution 
\cite{henkel2008non}, is assumed to scale
according to 
\begin{equation}
P_{\mathrm{wait}}(t; \delta \rho) \simeq \lambda^{-\epsilon_\parallel} 
\tilde{P}_{\mathrm{wait}} (\lambda^{-\nu_\parallel} t; \lambda \delta \rho)\,,
\end{equation}
which implies that
\begin{equation}
P_{\mathrm{wait}}(t;0) \sim t^{-\epsilon_{\parallel}/\nu_\parallel}\,,
\qquad\quad
\epsilon_{\parallel}/\nu_\parallel\in (1,2)\,.
\label{P_wait_epsilon_nu}
\end{equation}
In order for $P_{\mathrm{wait}}=P_{\mathrm{wait}}(t;0)$
to be normalizable one has that 
$\epsilon_{\parallel}/\nu_\parallel>1$. A diverging expectation
value $\langle\tau_{\mathrm{wait}}\rangle\to\infty$
at criticality needs conversely that
$\epsilon_{\parallel}/\nu_\parallel<2$.
We note that one could also use the scaling relation
$\epsilon_{\parallel}=2\nu_{\parallel}-\beta$.

In Fig.~\ref{fig:AAS_Waiting_1D_2D} the results for the
mean waiting time $\langle\tau_{\mathrm{wait}}\rangle$
and for the waiting-time distribution $P_{\mathrm{wait}}$ 
are presented for both $D\!=\!1$ and $D\!=\!2$. One
confirms that $\langle\tau_{\mathrm{wait}}\rangle$
scales inversely to the density of active sites as
a function of $p-p_c$, the distance to criticality.
The waiting-time distribution $P_{\mathrm{wait}}$
obeys (\ref{P_wait_epsilon_nu}) over several orders
of magnitude, with the range being somewhat restricted
in $D\!=\!2$ due to finite size effects. Overall
one finds, as listed in in the Tables \ref{tab_exponents_1D} 
and \ref{tab_exponents_2D}, good agreement with
directed percolation.

Experimentally, the waiting-time distribution is often directly
accessible, e.g.\ when measuring neural activity in terms of
individual spikes. Alternatively one may subdivide the experimental
timeline into a a series of time slices and define neural avalanches 
by counting the number of spikes observed in individual time slices
\cite{beggs2003neuronal,hesse2014self}. Obtaining neuronal 
avalanches in this way is however plagued by subsampling and
other issues \cite{priesemann2013neuronal}, which do not
affect local quantities such as the waiting-time distribution. 
Observables, as the waiting-time distribution, that do not
require to define neuronal avalanches in first place can hence 
be regarded as avalanche-free probes to absorbing phase transitions.

\subsection{The influence of metastable states\label{sect_metatstable}}

The AB-sublattice symmetry present in bipartite models 
is generically broken for synchronous sandpile models 
when nearest-neighbor toppling rules are applied, as 
discussed in Sect.~\ref{sect_AB}. The process of 
stochastically breaking the AB-symmetry may be fast or 
slow, as illustrated in Figs.~\ref{sublatticeTransients_2D} 
and \ref{sublatticeTransients_1D} respectively for the 
Manna- and for the AAS model. In particular
in $D\!=\!2$ we have the situation that the 
AB-symmetric state is transiently stable even close 
to criticality, see Fig.~\ref{phaseDiagram}.

The situation is non-trivial, as the state for
which all active sites are alternatingly on the A
and the B sublattice, the asymmetric state,
is not dynamically stable on short time scales.
A small perturbation in the form of a small addition
of active sites to the empty sublattice 
will quickly increase in magnitude,
until an equilibrated but fluctuating state is reached, 
as shown in Fig.~\ref{sublatticeTransients_1D}. The
sublattice breaking state is then recovered once 
a fluctuation in the sublattice density of active sites 
is large enough to empty one of the two sublattices 
altogether, which happens however only for prolonged 
times. This observation suggests that the AB-sublattice 
symmetry is broken stochastically and not, as we also
show in Sect.~\ref{sect_MFT_AB}, by a driving molecular
field.

The presence of metastable states is not expected 
to alter the critical exponents, it may however 
affect, as we speculate here, the width of the scaling 
regime, with the reason being that the state for which 
scaling is evaluated, the symmetry-broken state, is 
unstable for short times. 
This presumption could explain the non-monotonic 
behavior of $\Delta g$ in Fig.~\ref{AAS_active_g_Delta_2D} 
and that the deviations from a pure power-law scaling 
start in two dimensions at smaller $p-p_c$ than 
for $D\!=\!1$, compare Fig.~\ref{AAS_active_g_Delta_1D}
and \ref{AAS_active_g_Delta_2D}.

\section{Mean-field theory}
\label{sect_mean_field_theory}

The evolution of the densities $\rho_0$, $\rho_1$ and $\rho_a$ 
of empty, singly-occupied and active sites is 
governed on a molecular-field level by the master equation
\begin{eqnarray}
\label{ME_p0}
\rho_0|_{t+1} &=& (1-\rho_1)q_0 \\
\rho_1|_{t+1} &=& \rho_1q_0 + (1-\rho_1)q_1\,,
\label{ME_p1}
\end{eqnarray}
where $q_0$ and $q_1$ denote the probability to 
receive zero or one sandcorns from nearest-neighbor 
toppling sites. We have used here that synchronous
updating implies that a site is empty before receiving 
grains from other sites either because it was empty 
to start with, or because it was active. As described 
by (\ref{ME_p0}), a site will be empty if it was
not singly-occupied, the factor $1-\rho_1$, and if it 
did not receive grains, the factor $q_0$. Note that
(\ref{ME_p0}) and (\ref{ME_p1}) are not explicitly
dependent on the density $\rho_a=1-\rho_0-\rho_1$ of 
active sites. Once equilibrated, one has
for all $q_0,q_1\in[0,1]$
\begin{equation}
\rho_0=\frac{(1-q_0)q_0}{1-q_0+q_1},
\quad\quad
\rho_1=\frac{q_1}{1-q_0+q_1},
\quad\quad
\rho_a=\frac{(1-q_0)^2}{1-q_0+q_1}\,.
\label{rho_01_steadyState}
\end{equation}
For the evaluation of $q_0$ and $q_1$ for a 
$D$-dimensional hypercubic lattice we define with
\begin{equation}
\tilde{q}_0 = \left(1-\frac{1}{2D}\right)^2(1-p)
+\left(1-\frac{1}{2D}\right)^3p
\label{tilde_q_0}
\end{equation}
the probability that a selected neighboring site 
of a toppling site receives zero grains. In analogy 
we have
\begin{equation}
\tilde{q}_1 = \frac{1}{2D}
\left[2 \left(1-\frac{1}{2D}\right)(1-p)
+3\left(1-\frac{1}{2D}\right)^2p \right]
\label{tilde_q_1}
\end{equation}
that a selected neighboring site of a toppling sites
receives exactly one grain. One then has
\begin{equation}
q_0 = \big(1-\rho_a + \rho_a\tilde{q}_0\big)^{2D},
\quad\quad
q_1 = 2D\rho_a\tilde{q}_1\big(1-\rho_a + \rho_a\tilde{q}_0\big)^{2D-1}\,,
\label{q_01_2D}
\end{equation}
which one needs to substitute into (\ref{rho_01_steadyState}).
Expanding the result in $\rho_a$ one finds
\begin{equation}
p_c = \left\{
\begin{array}{rl}
0.3246 &(D\!=\!1) \\
0.1481 &(D\!=\!2) 
\end{array}
\right.,
\qquad\quad
\lim_{D\to\infty} p_c=0\,.
\label{p_c_D_1_2}
\end{equation}
In infinite dimensions exactly two grains are distributed
during an toppling event. The average number of grains 
per site is
\begin{equation}
g = \rho_1 + \rho_a(2+p)
= \frac{q_1+(1-q_0)^2(2+p)}{1-q_0+q_1}\,,
\label{g_MF}
\end{equation}
where $2+p$ is the average number of grains distributed
by an active site. The mean-field exponents are 
$\beta=\tilde{\beta}=1$ and $\sigma=\tilde{\sigma}=2$.

\subsection{Mean-field theory for two sublattices\label{sect_MFT_AB}}

Eqs.~(\ref{ME_p0}), (\ref{ME_p1}) and
(\ref{q_01_2D}) can be generalized to the
case of two sublattices. Iterating the resulting
master equations one finds that the AB-sublattice 
symmetry is kept, when present, and restored 
when starting with asymmetric initial conditions,
which implies that the critical value $p_c$ 
for toppling three and not two grains remains 
unaffected by the presence of two distinct sublattices
and that the loss of the AB-sublattice symmetry observed 
numerically in Figs.~\ref{sublatticeTransients_2D} 
and \ref{sublatticeTransients_1D} is the result of 
a stochastic process, and not due to a driving molecular 
field. For a generalized contact process it has been 
shown on the other hand that the sublattice 
symmetry can be broken already on a mean-field level
\cite{de2011contact}. Both states, broken and symmetric, 
exist in this case in specific regions in phase space
that are connected by a second-order phase transition, 
except for a limiting case where the transition is discontinuous.
This is however not the case for the AAS model.

\section{Discussion}

Previously, in Sect.~\ref{sect_metatstable}, we
argued that the observed deviations from power-law scaling,
which are particularly prominent in two dimensions
and for the scaling of the density of grains, may be
induced by the presence of a metastable state. A
range of alternatives is worth discussing.

\subsection{Logarithmic corrections}

Multiplicative logarithmic corrections are generically
present in marginal scenarios \cite{wegner1973logarithmic}, 
in particular at the upper critical dimension 
\cite{lubeck1998logarithmic} and at bicritical points, 
namely when a transition changes from second to first 
order \cite{kenna2006scaling}. For directed
percolation and for the Manna universality class, 
which includes the conserved lattice gas 
\cite{rossi2000universality}, the upper critical
dimension is four \cite{lubeck2004universal,henkel2008non}, 
which implies that the deviation from power-law scaling
present in Fig.~\ref{AAS_active_g_Delta_2D} should 
not be due to a marginal scenario.

\subsection{Two parameter scaling\label{sect_two_par_scaling}}

The density of active sites $\rho_a$ is determined
for the Manna model by the average number of grains 
per site, which acts, being a conserved quantity, as
an external control parameter. For the AAS model 
two quantities, $\rho_a$ and the grain density 
$ g$, are determined by the external
control parameter, namely $p$. We find that the grain 
density is coupled to $\rho_a$, as both $\langle \rho_a \rangle$ and 
$\langle g\rangle-g_c$ show critical scaling, where 
$g_c$ is the critical grain density. In such a
situation it is possible that the scaling flow 
is determined not by one, but by two running parameters, 
becoming hence two-dimensional. Two-parameter scaling 
is not very frequent, but known to be present for 
polymers in $\theta$ solvents \cite{colby1990two}, for 
weak topological insulators \cite{mong2012quantum}, and 
for integer quantum-Hall states \cite{werner2015anderson}. 
Our findings point towards $\gamma'=\tilde{\gamma}'$, 
which would indicate that only a single scale determines 
the renormalization flow.
However, we cannot rule out residual differences between
$\beta$  and $\tilde{\beta}$. The deviation from power-law 
scaling observed in Fig.~\ref{AAS_active_g_Delta_2D},
could be interpreted in this case as a crossover phenomenon.

\subsection{Conserving vs.\ dissipative avalanches}

There are two types of avalanches in locally
conserving sandpile models on lattices with
open boundary conditions. Dissipative avalanches
are those reaching the boundary, where grains are 
lost, whereas conserving avalanches stop before 
reaching the boundary. Dissipative and conserving 
avalanches differ qualitatively with respect to their 
statistical properties, with dissipative avalanches 
showing clean power-law scaling \cite{drossel2000scaling}.
Conserving avalanches are characterized in contrast 
by substantial multiplicative logarithmic corrections, 
which are furthermore strongly size dependent 
\cite{dickman2003avalanche}. For this scenario to
explain the deviation of power-law scaling
present in Fig.~\ref{AAS_active_g_Delta_2D},
the internal avalanches of the AAS model with periodic
boundary conditions would need to behave akin to the
conserving avalanches of abelian sandpile models with
open boundaries. The rational for this putative 
equivalence is however presently not evident.

\section{Conclusions}

Surprisingly complex physics is found for a
new dynamical systems model, the autonomously 
adapting sandpile (AAS) model. While energy, 
respectively sand is locally non conserved,
an absorbing phase transition within the directed 
percolation universality class is observed. For
the AAS model the density of grains attains a
self-organized value.
Pointing out that synchronous update 
rules on bipartite lattices allow the AB-sublattice
symmetry to be broken stochastically, we argue that
the metastable sublattice-symmetric state leaves 
its imprint on the scaling behavior, in particular
in two dimensions, where the sublattice symmetric
state has a prolonged lifetime even in the vicinity to 
criticality. The proposed model is well suited
for modeling critical brain dynamics, as it mimics
the dynamics of integrate and fire neurons. Open
issues regard the lifetime of the metastable AB-symmetric
state and the behavior on non-bipartite lattices
and graphs, respectively of update rules that do not 
conserve the sublattice symmetry. 

\section*{Acknowledgments}

We thank Peter Kopietz, Malte Henkel and 
Dimitrije Markovic for discussions.

\section*{References}

\bibliographystyle{iopart-num}

\begin{thebibliography}{10}
\expandafter\ifx\csname url\endcsname\relax
  \def\url#1{{\tt #1}}\fi
\expandafter\ifx\csname urlprefix\endcsname\relax\def\urlprefix{URL }\fi
\providecommand{\eprint}[2][]{\url{#2}}

\bibitem{hinrichsen2000non}
Hinrichsen H 2000 {\em Advances in Physics\/} {\bf 49} 815--958

\bibitem{kockelkoren2003absorbing}
Kockelkoren J and Chat{\'e} H 2003 {\em Physical Review Letters\/} {\bf 90}
  125701

\bibitem{henkel2004non}
Henkel M and Hinrichsen H 2004 {\em Journal of Physics A: Mathematical and
  General\/} {\bf 37} R117

\bibitem{rossi2000universality}
Rossi M, Pastor-Satorras R and Vespignani A 2000 {\em Physical Review
  Letters\/} {\bf 85} 1803

\bibitem{manna1991two}
Manna S 1991 {\em Journal of Physics A: Mathematical and General\/} {\bf 24}
  L363

\bibitem{beggs2003neuronal}
Beggs J~M and Plenz D 2003 {\em Journal of Neuroscience\/} {\bf 23}
  11167--11177

\bibitem{priesemann2013neuronal}
Priesemann V, Valderrama M, Wibral M and Le~Van~Quyen M 2013 {\em PLoS
  Computational Biology\/} {\bf 9} e1002985

\bibitem{chialvo2004critical}
Chialvo D~R 2004 {\em Physica A: Statistical Mechanics and its Applications\/}
  {\bf 340} 756--765

\bibitem{markovic2014power}
Markovi{\'c} D and Gros C 2014 {\em Physics Reports\/} {\bf 536} 41--74

\bibitem{gros2013observing}
Gros C and Markovi{\'c} D 2013 Observing scale-invariance in non-critical
  dynamical systems {\em AIP Conference Proceedings\/} vol 1510 (AIP) pp 44--53

\bibitem{markovic2013criticality}
Markovi{\'c} D, Gros C and Schuelein A 2013 {\em Chaos: An Interdisciplinary
  Journal of Nonlinear Science\/} {\bf 23} 013106

\bibitem{markovic2009vertex}
Markovic D and Gros C 2009 {\em New Journal of Physics\/} {\bf 11} 073002

\bibitem{bak1998nature}
Bak P 1998 {\em How Nature Works: The Science of Self-Organized Criticality
  Copernicus, New York, 1996; HJ Jensen, Self-Organized Criticality\/}
  (Cambridge University Press, Cambridge)

\bibitem{bak1987self}
Bak P, Tang C and Wiesenfeld K 1987 {\em Physical Review Letters\/} {\bf 59}
  381

\bibitem{levina2017subsampling}
Levina A and Priesemann V 2017 {\em Nature Communications\/} {\bf 8} 15140

\bibitem{drossel2000scaling}
Drossel B 2000 {\em Physical Review E\/} {\bf 61} R2168

\bibitem{dickman2003avalanche}
Dickman R and Campelo J 2003 {\em Physical Review E\/} {\bf 67} 066111

\bibitem{vespignani1998driving}
Vespignani A, Dickman R, Mu{\~n}oz M~A and Zapperi S 1998 {\em Physical Review
  Letters\/} {\bf 81} 5676

\bibitem{gros2015complex}
Gros C 2015 {\em Complex and adaptive dynamical systems: A primer\/} (Springer)

\bibitem{basu2013absorbing}
Basu U, Basu M and Mohanty P 2013 {\em The European Physical Journal B\/} {\bf
  86} 236

\bibitem{henkel2008non}
Henkel M, Hinrichsen H, L{\"u}beck S and Pleimling M 2008 {\em Non-equilibrium
  phase transitions\/} vol~1 (Springer)

\bibitem{lubeck2001scaling}
L{\"u}beck S 2001 {\em Physical Review E\/} {\bf 64} 016123

\bibitem{lee2014critical}
Lee S~B 2014 {\em Physical Review E\/} {\bf 89} 062133

\bibitem{hohenberg1977theory}
Hohenberg P~C and Halperin B~I 1977 {\em Reviews of Modern Physics\/} {\bf 49}
  435

\bibitem{lubeck2003universal}
L{\"u}beck S and Heger P 2003 {\em Physical Review E\/} {\bf 68} 056102

\bibitem{daga2018universality}
Daga B and Ray P 2018 {\em arXiv preprint arXiv:1812.10704\/}

\bibitem{lubeck2004universal}
L{\"u}beck S 2004 {\em International Journal of Modern Physics B\/} {\bf 18}
  3977--4118

\bibitem{basu2012fixed}
Basu M, Basu U, Bondyopadhyay S, Mohanty P and Hinrichsen H 2012 {\em Physical
  review letters\/} {\bf 109} 015702

\bibitem{HINRICHSEN20061}
Hinrichsen H 2006 {\em Physica A: Statistical Mechanics and its Applications\/}
  {\bf 369} 1 -- 28 ISSN 0378-4371 fundamental Problems in Statistical Physics

\bibitem{mohanty2002generic}
Mohanty P and Dhar D 2002 {\em Physical Review Letters\/} {\bf 89} 104303

\bibitem{lubeck2002scaling}
L{\"u}beck S 2002 {\em Physical Review E\/} {\bf 65} 046150

\bibitem{lubeck2002universal}
L{\"u}beck S and Willmann R 2002 {\em Journal of Physics A: Mathematical and
  General\/} {\bf 35} 10205

\bibitem{hesse2014self}
Hesse J and Gross T 2014 {\em Frontiers in systems neuroscience\/} {\bf 8} 166

\bibitem{de2011contact}
de~Oliveira M~M and Dickman R 2011 {\em Physical Review E\/} {\bf 84} 011125

\bibitem{wegner1973logarithmic}
Wegner F~J and Riedel E~K 1973 {\em Physical Review B\/} {\bf 7} 248

\bibitem{lubeck1998logarithmic}
L{\"u}beck S 1998 {\em Physical Review E\/} {\bf 58} 2957

\bibitem{kenna2006scaling}
Kenna R, Johnston D and Janke W 2006 {\em Physical Review Letters\/} {\bf 96}
  115701

\bibitem{colby1990two}
Colby R~H and Rubinstein M 1990 {\em Macromolecules\/} {\bf 23} 2753--2757

\bibitem{mong2012quantum}
Mong R~S, Bardarson J~H and Moore J~E 2012 {\em Physical Review Letters\/} {\bf
  108} 076804

\bibitem{werner2015anderson}
Werner M~A, Brataas A, Von~Oppen F and Zar{\'a}nd G 2015 {\em Physical Review
  B\/} {\bf 91} 125418

\end{thebibliography}

\providecommand{\newblock}{}



\end{document}